\documentclass[%
reprint,
superscriptaddress,
amsmath,amssymb,
aps,
table]{revtex4-2}
\usepackage{comment}
\usepackage{svg}
\usepackage{float}
\usepackage{mathtools}
\usepackage{siunitx}
\usepackage{graphicx}% Include figure files
\usepackage{dcolumn}% Align table columns on decimal point
\usepackage{bm}% bold math
\usepackage{soul} % allows striking through text for editing purposes, can delete later
\usepackage{hyperref}
\hypersetup{colorlinks = true,
            linkcolor = blue,
            urlcolor  = blue,
            citecolor = blue,
            anchorcolor = blue}

\usepackage{verbatim}

\usepackage{xspace}
\newcommand{\PI}{Phase~I\xspace}
\newcommand{\PII}{Phase~II\xspace}
\newcommand{\PIIa}{Phase~IIa\xspace}
\newcommand{\PIIb}{Phase~IIb\xspace}
\newcommand{\PIIc}{Phase~IIc\xspace}
\newcommand{\PIId}{Phase~IId\xspace}

\newcommand{\PIIcd}{Phase~IIc/d\xspace}
\newcommand{\PIIab}{Phase~IIa/b\xspace}

\newcommand{\SC}{$S_{c}$\xspace}
\newcommand{\NC}{$N_{c0}$\xspace}

\newcommand{\NA}{$N_{a}$\xspace}
\newcommand{\SRE}{$E$\xspace}

\newcommand{\PIIFreqCovMHz}{\SI{550}{\MHz}\xspace}
\newcommand{\PIIcdFreqCovMHz}{\SI{413}{\MHz}\xspace}
\newcommand{\PIIcFreqCovMHz}{184\xspace}
\newcommand{\PIIdFreqCovMHz}{229\xspace}

\newcommand{\numRFI}{12\xspace}
\newcommand{\TMC}{\SI{60}{mK}\xspace}
\newcommand{\TMCerr}{$60\pm$\SI{1}{mK}\xspace}
\newcommand{\TVTSerr}{$290\pm$\SI{10}{mK}}
\newcommand{\PIIJointAgg}{2.86$\times$$|g_{\gamma}^{\text{KSVZ}}|$\xspace}
\newcommand{\PIIcJointAgg}{2.96$\times$$|g_{\gamma}^{\text{KSVZ}}|$\xspace}
\newcommand{\PIIdJointAgg}{2.75$\times$$|g_{\gamma}^{\text{KSVZ}}|$\xspace}

\begin{document}
%\detailtexcount{main}

\preprint{APS/123-QED}

\title{Dark Matter Axion Search with HAYSTAC Phase~II}% Force line breaks with \\
%\thanks{tentative title; suggestions are welcome.}%
\newcommand{\Location}{\affiliation{Location}}
\newcommand{\Yale}{\affiliation{Department of Physics, Yale University, New Haven, Connecticut 06520, USA}}
\newcommand{\YaleApplied}{\affiliation{Department of Applied Physics, Yale University, New Haven, Connecticut 06520, USA}}
\newcommand{\WrightLab}{\affiliation{Wright Laboratory, Department of Physics, Yale University, New Haven, Connecticut 06520, USA}}
\newcommand{\Cal}{\affiliation{Department of Nuclear Engineering, University of California Berkeley, California 94720, USA}}
\newcommand{\JILA}{\affiliation{JILA, National Institute of Standards and Technology and the University of Colorado, Boulder, Colorado 80309, USA}}
\newcommand{\Colorado}{\affiliation{Department of Physics, University of Colorado, Boulder, Colorado 80309, USA}}
\newcommand{\Hopkins}{\affiliation{Department of Physics and Astronomy, The Johns Hopkins University, Baltimore, MD, 21218}}
\newcommand{\NIST}{\affiliation{National Institute of Standards and Technology, Boulder, Colorado 80305, USA}}

\author{Xiran~Bai}\Yale\WrightLab
\author{M.J.~Jewell}\Yale\WrightLab
\author{J.~Echevers}\Cal

\author{K.~van~Bibber}\Cal
%\author{S.B.~Cahn}\Yale\WrightLab
\author{A.~Droster}\Cal
\author{Maryam H.~Esmat}\Hopkins
\author{Sumita~Ghosh}\altaffiliation{now at Massachusetts Institute of Technology}\YaleApplied\WrightLab
\author{Eleanor~Graham}\Yale\WrightLab
\author{H.~Jackson}\Cal
\author{Claire~Laffan}\Yale\WrightLab
\author{S.K.~Lamoreaux}\Yale\WrightLab
\author{A.F.~Leder}\altaffiliation{now at Los Alamos National Laboratory}\Cal
\author{K.W.~Lehnert}\Yale\WrightLab
\author{S.M.~Lewis}\altaffiliation{now at Wellesley College}\Cal
\author{R.H.~Maruyama}\Yale\WrightLab
\author{R.D.~Nath}\Cal
\author{N.M.~Rapidis}\altaffiliation{now at Stanford University}\Cal
\author{E.P.~Ruddy}\Colorado\JILA
\author{M.~Silva-Feaver}\Yale\WrightLab
\author{M.~Simanovskaia}\altaffiliation{now at Stanford University}\Cal
\author{Sukhman~Singh}\Yale\WrightLab
\author{D.H.~Speller}\Hopkins
\author{Sabrina Zacarias}\Yale\WrightLab
\author{Yuqi~Zhu}\altaffiliation{now at Stanford University}\Yale\WrightLab

\collaboration{HAYSTAC Collaboration}
\date{\today}% It is always \today, today,

\begin{abstract}
%TC:ignore
This Letter reports new results from the HAYSTAC experiment's search for dark matter axions in our galactic halo. It represents the widest search to date that utilizes squeezing to realize sub-quantum limited noise. The new results cover 1.71~$\mu$eV of newly scanned parameter space in the mass ranges 17.28--18.44~$\mu$eV and 18.71--19.46 $\mu$eV. No statistically significant evidence of an axion signal was observed, excluding couplings $|g_\gamma|\geq$~\PIIdJointAgg and  $|g_\gamma|\geq$~\PIIcJointAgg at the 90$\%$ confidence level over the respective region. By combining this data with previously published results using HAYSTAC's squeezed state receiver, a total of 2.27~$\mu$eV of parameter space has now been scanned between 16.96--19.46 $\mu$eV, excluding $|g_\gamma|\geq$~\PIIJointAgg at the 90$\%$ confidence level. These results demonstrate the squeezed state receiver's ability to probe axion models over a significant mass range while achieving a scan rate enhancement relative to a quantum-limited experiment. 

\begin{comment}
\begin{description}
\item[Usage]
Secondary publications and information retrieval purposes.
\item[Structure]
You may use the \texttt{description} environment to structure your abstract;
use the optional argument of the \verb+\item+ command to give the category of each item. 
\end{description}
\end{comment}

\end{abstract}

%TC:ignore
\maketitle
%TC:endignore
%\tableofcontents

%\section{Introduction\label{sec:intro}}
 \textit{Introduction}---One of the most pressing questions in physics is the nature of dark matter, for which the quantum chromodynamics (QCD) axion offers a compelling solution. The axion arose from the theory of Peccei and Quinn (PQ) to explain the absence of CP violation in QCD \cite{peccei1977CP,peccei1977CP2, Weinberg:1977ma,Abbott:1982af,Preskill:1982cy,dine1983harmless}. While the mass of these QCD axions remains unknown, in the case that PQ symmetry breaking occurs after inflation, axions within the mass range $m_a\sim$ 1--500 $\mu eV$ are favored \cite{gorghetto2019mass,klaer2017dark,buschmann2020motivation,saikawa2024spectrumglobalstringnetworks,O_Hare_2022,buschmann2022mesh}. 
 
 To date the most sensitive probes for QCD axions in this range are those using axion haloscopes~\cite{RYBKA2024116481,Sikivie:1983ip_halotheory,Sikivie:1985yu_halotheory}. In a haloscope experiment, a magnetic field is used to convert the oscillating axion field into an electric field that oscillates at $\nu_a \approx m_a c^2/h$. A tunable cavity is used to resonantly enhance the conversion power, which is maximized when its resonant frequency ($\nu_c$) matches that of the axion ($\nu_a$). The signal power, in natural units, is given by 
 
\begin{equation*}
%TC:ignore
     P_{ax} = \left(\frac{g^{2}_{\gamma}\alpha^2\rho_a}{\pi^2 \Lambda^4}\right) \omega_c B^{2}_{0} V C_{\text{mnl}} Q_{L} \frac{\beta}{1+\beta}\frac{1}{1 +(2{\delta_\nu}/\Delta\nu_{c})^2}
 %TC:endignore
\end{equation*} 

\noindent The leading parentheses contain physical constants and parameters predicted by dark matter axion models, where $g_\gamma$ is a dimensionless parameter defining the axion-photon coupling strength which is predicted to be -0.97 (0.36) by the benchmark KSVZ (DFSZ) axion models~\cite{kim1979KSVZ,shifman1980KSVZ2,dine1981DFSZ,zhit1980DFSZ2}, $\alpha$ is the fine-structure constant, $\rho_{a}$ is the local dark matter density taken here as 0.45 GeV/cm$^3$, and $\Lambda$ defines the zero-temperature QCD topological susceptibility taken here as \SI{77.6}{MeV}~\cite{olive_review_2016}. The remaining parameters are properties of the detector, where $\omega_c = 2\pi\nu_c$, $B_0$ is the strength of the magnetic field, $V$ is the unfilled volume of the cavity, and $C_{\text{mnl}}$ is the normalized form factor which quantifies the overlap of the B-field with the cavity mode and signifies the coupling of the axion to a specific cavity mode~\cite{Sikivie:2020zpn}. The lowest order of $\text{TM}_{0n0}$ mode is commonly used as it has the largest overlap between the external B-field and the internal E-field. Finally, the last set of terms describes the scaling from detuning of the axion signal from the cavity's resonance, $\delta_\nu = \nu_a - \nu_c$, which follows a Lorentzian with linewidth given by the loaded quality factor ($Q_L$) as $\Delta\nu_c =  \nu_c/Q_L$.  The loaded $Q$ is related to the unloaded $Q$ ($Q_0$) by the coupling coefficient $\beta$ as $Q_L=Q_0/(\beta+1)$.

A major challenge in axion searches comes from the standard quantum limit~(SQL) on the noise added by phase-insensitive linear amplifiers typically used by haloscopes~\cite{Haus:1282407,Caves:1982zz}. To address this, the Haloscope at Yale Sensitive To Axion Cold Dark Matter~(HAYSTAC) utilizes quantum squeezing to reach noise levels below the SQL~\cite{backes2021quantum,HAYSTAC_2023MJ,Kelly_thesis, palken2020thesis}. This is realized by coupling the cavity to a squeezed state receiver~(SSR) consisting of two Josephson parametric amplifiers~(JPAs) operating as phase-sensitive amplifiers, reducing the noise in one of the measured quadratures below the SQL. This does not improve the peak sensitivity at $|\delta_\nu|=0$, but instead results in an increased visible bandwidth by reducing noise off resonance~($|\delta_\nu|>0$).  This allows larger $\beta$'s to be used to improve the scan rate over what is achievable with a quantum-limited search~\cite{yamamoto2008flux,malnou2019squeezed,Malnou:2017udw}.

%\st{This reduces the noise at $|\delta_\nu|>0$, allowing for a scan rate enhancement over quantum-limited searches when increasing the cavity bandwidth with stronger couplings to the cavity mode}~\cite{yamamoto2008flux,malnou2019squeezed,Malnou:2017udw}.

Following the successful conclusion of HAYSTAC's \PI in 2017~\cite{brubaker2017first,zhong2018results}, \PII operation began in September 2019 and ended in August 2024 with operations divided into four subphases (Phase~IIa-d), which in total covered \PIIFreqCovMHz of parameter space as summarized in Table \ref{tab:summary}. Since the initial demonstration of squeezing in \PIIa~\cite{backes2021quantum}, HAYSTAC has continued operating the SSR, with additional results from \PIIb~\cite{HAYSTAC_2023MJ}. The results reported in this Letter (\PIIcd) show new data taken between 4.178--4.459 GHz and 4.523--4.707 GHz, covering a range of \PIIcdFreqCovMHz. With these results, HAYSTAC has demonstrated the capability of the SSR to enhance QCD axion searches over a significant range of masses.

\textit{Experimental Details}---The HAYSTAC experiment~\cite{Kenany2017design, brubaker2017first, zhong2018results, backes2021quantum, HAYSTAC_2023MJ}, located at Yale University's Wright Laboratory, consists of a tunable microwave cavity installed inside a magnet aligned in $\hat{z}$ and operated at \SI{8}{T}  which is coupled to an SSR. To reduce the noise temperature, the cavity and the receiver chain are operated in a dilution fridge at \TMC. The cavity consists of a copper-plated stainless steel cylinder and a single off-axis tuning rod made of the same material, leaving an unfilled volume of $V = \SI{1.545}{L}$. Simulations of the cavity in CST~\cite{CST} are used to determine the form factor for different mode frequencies.  These results are validated with bead perturbations~\cite{rapidis_characterization_2019} and find an average $C_{010}$ of $0.43\pm0.01$ over the range scanned here.  A vector network analyzer~(VNA) is used to extract $\beta$ and $Q_{L}$ from the cavity's response each time the cavity is tuned.  To remain within $5\%$ of the maximum scan rate of the SSR the cavity's $\beta$ was maintained between 6.4--11.1, with an average of $\beta=8.5\pm0.3$.   Combined with $Q_{L}$ this gives an average $Q_0$ of  $44000 \pm 4000$, where both here and above the error bars represent measurement uncertainty.  

%The frequency dependent form factor is found via simulations of the cavity in CST~\cite{CST}, which are validated with bead perturbations~\cite{rapidis_characterization_2019}, and over the range scanned here, the average is $C_{010}=0.43\pm0.01$. 

\begin{figure}
    \centering
    \includegraphics[width=1\linewidth]{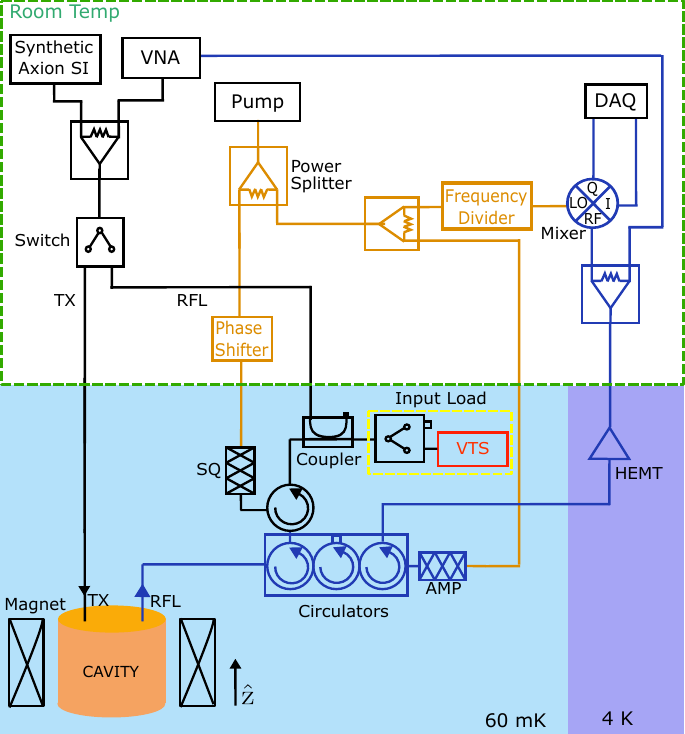}
    \caption{Simplified version of the HAYSTAC receiver described in~\cite{HAYSTAC_2023MJ}. In black are inputs to the system which can be sent to either the cavity's transmission~(TX) or reflection~(RFL) port. These are used for characterizing the system with a VNA and for injecting synthetic axion signals into the system. The orange lines indicate the path of the pump tone used by both the JPAs and the IQ mixer. Finally, shown in blue is the readout path which extracts and amplifies signals from the cavity for readout.}
    \label{fig:sys_diagram}
\end{figure}

As shown in Fig.~\ref{fig:sys_diagram} detection of the cavity field is achieved by coupling an antenna to the cavity. The signal is then amplified with a cryogenic receiver chain, the SSR, which consists of two JPAs. The first JPA~(SQ) prepares the vacuum noise (sourced from a \SI{50}{\ohm} terminator held at \SI{60}{mK}) in a squeezed state, reducing the variance of the noise to below the vacuum level along one quadrature. The state is then reflected off of the cavity where it picks up cavity noise, which could contain an axion signal. The second JPA~(AMP) amplifies the state along the previously squeezed quadrature, and the output signal is fed into the subsequent amplification chain and recorded by the digitizer.  

New to \PIIcd is the injection of synthetic axion signals~(SI) whose spectral shape match the expected virialized axion line shape~(ALS) with linewidth $\delta_{\nu_{a}}$$\sim5$~kHz.  As detailed in~\cite{Zhu_2023} this can be accomplished by ``hopping" the frequency output by a signal generator between values sampled from the boosted Maxwell-Boltzmann distribution which defines the ALS~\cite{Turner_1990}. Furthermore, these signals are injected via the cavity's transmission port such that they are read out and amplified by the subsequent amplification chain in the same way as an axion signal.  This allows injections to be used as a validation of the receiver chain and analysis pipeline, but given the uncertainty in the transmission efficiency from the signal generator and through the cavity, the injections are not used as an independent calibration.  Instead, the amplitude relative to the noise, measured with the standard calibration routine, is used to calibrate the signals.  During \PIIcd, six such signals were injected corresponding to coupling strengths between $3.04-14.00 \times|g^{\text{KSVZ}}_{\gamma}|$.

%As such, these injections allow us to validate the receiver chain and analysis pipeline, and during \PIIcd six such signals were injected. 

\begin{figure}[!]
    \centering
    \includegraphics[width=\linewidth]{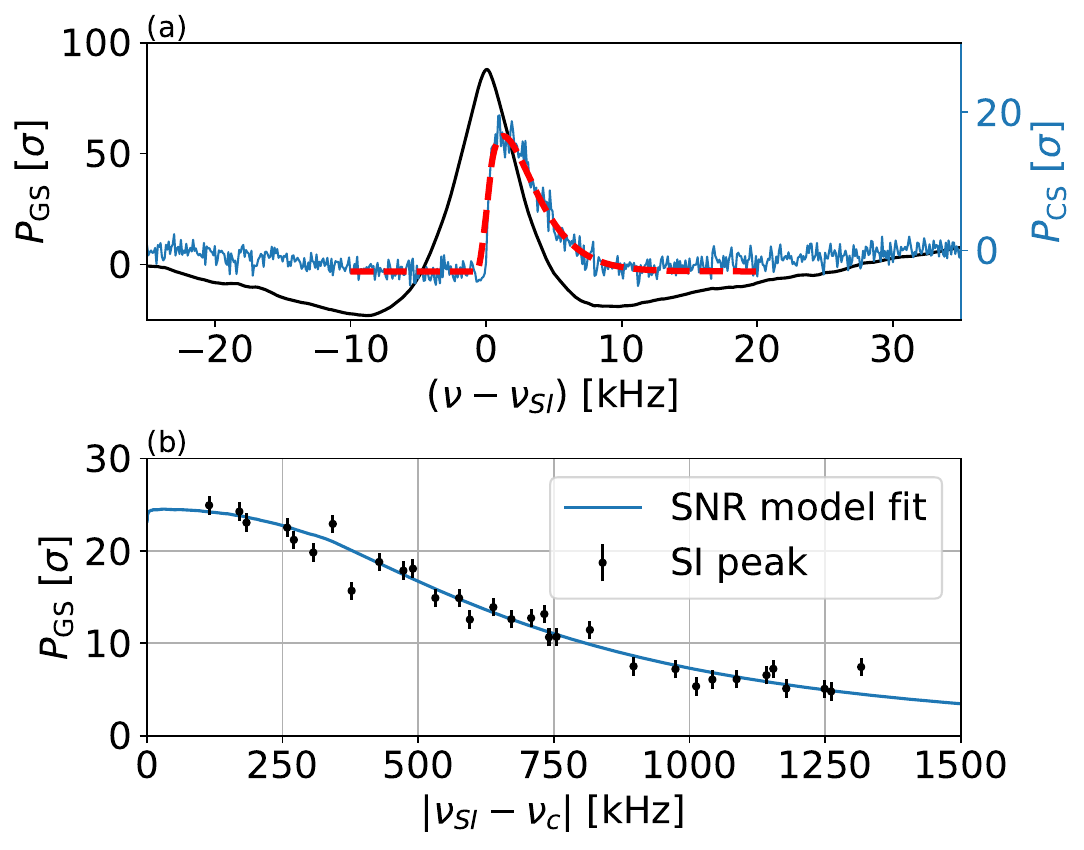}
    \caption{(a) Normalized power observed for a synthetic axion signal corresponding to an axion with $|g_{\gamma}|=14.00\times|g^{\text{KSVZ}}_{\gamma}|$ injected at \SI{4.459220}{GHz} during \PIId. Shown for the combined spectrum~(blue) and grand spectrum~(black) and compared to the expected ALS~(red dashed line). (b) When the PSD from each tuning step is processed separately, instead of being combined, the dependence of the injected signal's peak power on its detuning from the cavity's resonant frequency is seen to agree with the scaling predicted by the SNR model~(blue solid line).  Powers at detunings  $<$\SI{45}{\kHz} are removed as this is outside of the analysis band.}
    \label{fig:SI}
\end{figure}
%of the injected signal observed in the average PSD for each tuning step~(black points) is shown as a
%Note that the power in (b) is for each PSD before summing up to create a grand spectrum. Therefore, adding all peak powers in (b) results in a peak power of 97.3 $\sigma$ in (a).

During standard data taking, a fully automated script first tunes the cavity's resonance by rotating the tuning rod, producing a frequency step size of $\sim$\SI{80}{kHz}, and then sets the pump frequency of the JPAs to twice the cavity frequency measured by the VNA. To ensure optimal sensitivity, the phase shift between the two JPAs along with the JPA bias currents and pump powers are tuned to optimize the squeezing. At each tuning step, the digitizer records data in \SI{10}{ms} segments at a sampling rate $\geq$~\SI{5}{MHz} and computes the average power spectral density (PSD) over time $\tau$. During \PIIcd an integration time of 20 min was used, lower than the 60 min used in \PIIab to allow for a larger range of frequencies to be covered.

\begin{table}[]
\begin{tabular}{c|c|c|c|c|ccc|}
  & Op. {[}days{]} & $\Delta\nu$ {[}MHz{]} & $|g_{\gamma}/g_{\text{KSVZ}}|$  & Cand. & RFI & Def. & SI      \\ \hline
a & 106            & 73            & 2.05       & 32    & -   & 5    & -  \\
b & 53            & 64             & 1.96       & 37     & -   & 1    & -  \\
c & 72            & \PIIcFreqCovMHz           & 3.04        & 98    &7   & 8    & 3   \\
d & 111            & \PIIdFreqCovMHz          & 3.06        & 120   &5   & 16   & 3   
\end{tabular}    
\caption{Summary of HAYSTAC's \PII operations showing total operational days, frequency coverage, and the average $|g_{\gamma}|$ which would produce a 5.1$\sigma$ signal for the initial scans. Also shown are the total number of rescan candidates along with the individual contributions from the nonstatistical fluctuations (rf interference, deficits, and synthetic axion SI) described in the text.}
\label{tab:summary}
\end{table} 

Calibration of the noise performance follows a similar routine to that described in~\cite{backes2021quantum, HAYSTAC_2023MJ, Malnou:2017udw}, with the system noise expressed using single-quadrature spectral densities. The calibration starts with a standard Y-factor measurement with the JPAs detuned from the cavity to extract the added noise of the amplification chain, \NA, referred to the input of the AMP. This is achieved by comparing the PSD for two different input loads on either side of the switch. One thermalized to the mixing chamber at \TMCerr, and the other to a variable temperature stage~(VTS) held at \TVTSerr. This measurement is repeated approximately once per nine tuning steps during normal operations to capture potential variations with frequency and time.  Over the range covered in \PIIcd, \NA was observed to vary between $0.01<N_a<0.18$ quanta with an average value of $N_a=0.08\pm0.07$ quanta over the analysis band, with this and subsequent error bars representing systematic error. This agrees, within the uncertainty, to the specifications of the HEMT noise which is expected to be the dominant contribution in the idealized JPA model. The above measurement of \NA requires the transmission efficiency between the input load and the AMP. Following a similar routine to~\cite{malnou2019squeezed}, this is measured in two steps by varying which JPA is used as the main amplifier to decouple the efficiency between the input load and the SQ~($\lambda$) from the efficiency between the SQ and AMP~($\eta$).  To account for possible frequency-dependent variations, this measurement is repeated over the full range of frequencies covered in \PIIcd, in steps of \SI{3}{\MHz}, outside of normal operations.  Over this range, $\lambda$ is found to be generally frequency independent, with an average value of  $\lambda=0.67\pm0.05$, while $\eta$ was found to vary by 12$\%$ with an average value of $\eta=0.65\pm0.03$.  

Similar to \PIIab, the observed noise contribution of the cavity, \NC, is larger than would be expected for a cavity thermalized to the mixing chamber. This is quantified by periodically comparing the PSD with and without the cavity present. During \PIIcd, \NC exhibited repeatable variations with the TM$_{010}$ frequency, with \NC ranging between $0.43<\mbox{\SC}<1.20$ quanta and having an average of \NC$=0.71\pm0.15$ quanta over the full range of \PIIcd.  While the cause of this excess noise is still being investigated, it has previously been attributed to poor thermalization of the tuning rod but could also be the result of noise coupling through one of the cavity ports. 

The final step in characterizing the performance is a measurement of the noise reduction from squeezing. Following~\cite{malnou2019squeezed,backes2021quantum, HAYSTAC_2023MJ}, this is quantified by comparing the PSD taken with and without the SQ.  Measurements taken over the frequency range of \PIIcd,  with the JPAs tuned away from the cavity resonance, show off-resonant squeezing ranging between $3.5<S<4.0$~dB, in agreement with the expected performance given the measured values of $\eta$ and \NA.  However, the observed squeezing during normal operations, with the JPAs tuned to match the cavity frequency, was lower than expected from the off-resonant measurements.  It is suspected that the additional noise, with a similar shape to the cavity Lorentzian, is caused by mechanical vibrations of the cavity tuning rod or the antenna.  These vibrations cause the TM$_{010}$ mode to oscillate at a rate below $\sim$\SI{200}{\Hz}, resulting in dephasing of the squeezed state between the two JPAs relative to the optimal \SI{90}{\degree} phase difference.  Because this noise varies with both time and frequency, a measurement with and without the SQ is taken before each PSD to extract the squeezing as a function of detuning. To understand the impact on the sensitivity, this can be translated into a scan rate enhancement~(\SRE) relative to operation without squeezing~\cite{malnou2019squeezed}. In the absence of additional noise from vibrations, an enhancement between $1.8<\mbox{\SRE}<2.0$ would be expected given the off-resonant performance. However, taking into account the observed noise in \PIIcd, the average enhancement achieved was $\mbox{\SRE}=1.7$ with the enhancement varying between $1.3<\mbox{\SRE}<2.0$ over the full operating range.

\textit{Analysis and Results}---Data recorded by the digitizer is analyzed largely following the procedure described in~\cite{HAYSTAC_2023MJ,brubaker2017analysis}, which starts with a set of data quality cuts to remove PSDs that exhibit poor or anomalous behavior such as unstable gains, non-optimal squeezing, or degraded cavity performance. In total, these cuts remove $\sim$3$\%$ of the PSDs. The remaining PSDs are then processed with a Savitzky-Golay~(SG) filter to remove structures wider than expected from the ALS. This is achieved by dividing each PSD by the output of a Savitzky-Golay~(SG) filter and then subtracting the unitary mean. In the absence of a signal, the processed spectra are approximately Gaussian distributed with $\mu=0$ and $\sigma=1/\sqrt{\Delta_b\tau}$, where $\Delta_b$~=~\SI{100}{Hz} is the frequency resolution of the FFT. The spectra are then scaled by the maximum likelihood weights given by the signal-to-noise ratio~(SNR), aligned by their rf frequency, and summed to produce a single combined spectrum showing the observed power excess at each probed frequency. To optimize the sensitivity to an axion with the assumed line shape, the ALS weighted sum of adjacent bins is taken, and a correction to remove effects introduced by the SG filter is applied to obtain the final grand spectrum.

\begin{figure*}
    \centering
    \includegraphics[width=\linewidth]{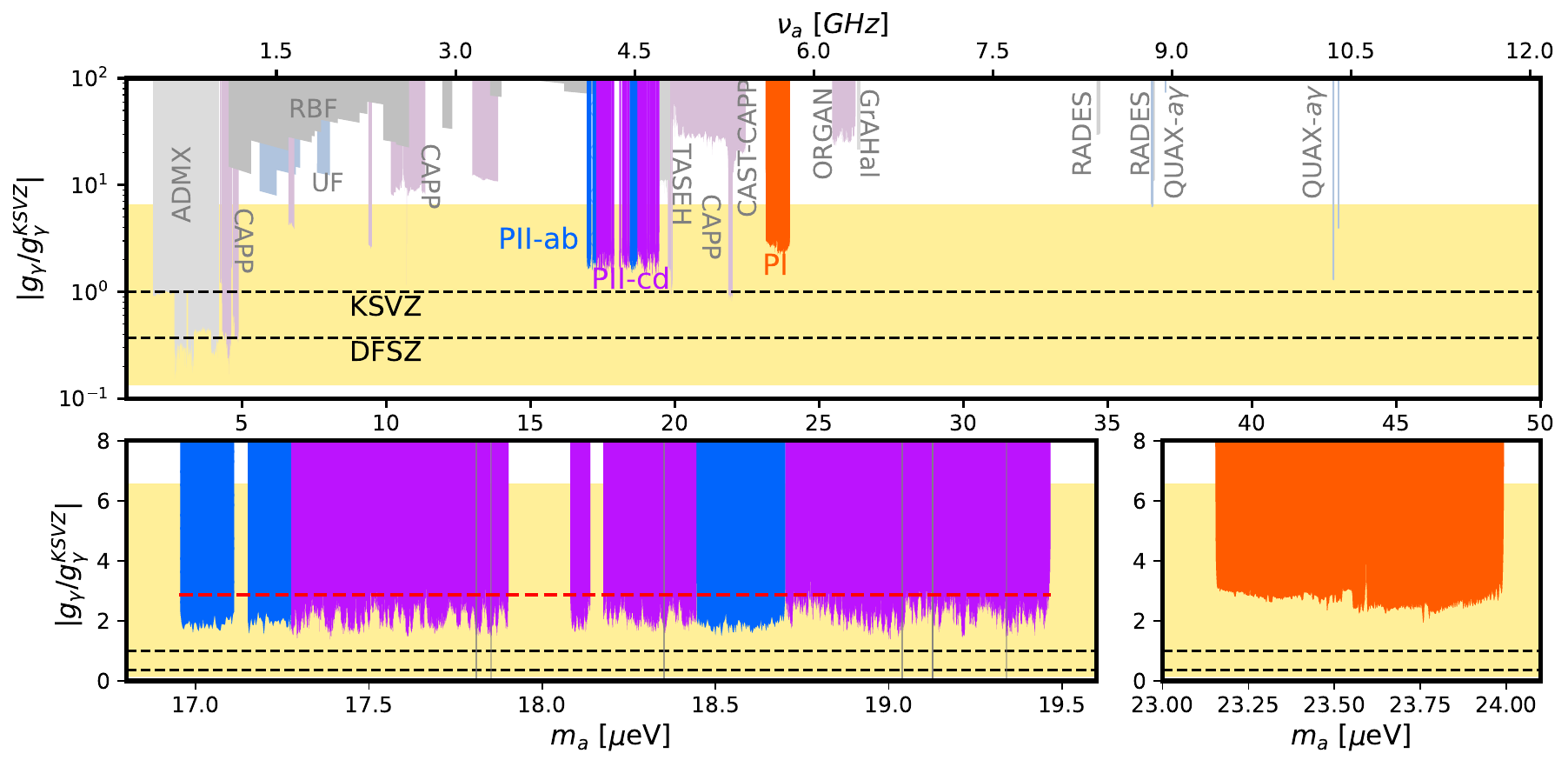}
    \caption{Top: Current exclusion on the axion-photon coupling in the 1--\SI{50}{\mu eV} mass range showing results from HAYSTAC alongside results from ADMX~\cite{ADMX2010, DuADMX:2018, BraineADX2020, ADMX2021, ADMX2024, ADMXSidecar2018, ADMXSidecar2021}, CAPP~\cite{CAPP1:2020, CAPP2:2020, CAPP3:2021, CAPP4:2022, CAPP4:2022_analysis, CAPP5:2023, CAPP6:2023, CAPP7:2023, CAPP8:2024, CAPPMAX:2024}, CAST-CAPP~\cite{CASTCAPP:2022}, GrAHal~\cite{GRAHAL:2021},  ORGAN~\cite{ORGAN:2024}, QUAX~\cite{QUAX:2019, QUAX:2021, QUAX:2022, QUAX:2023, QUAX:2024}, RADES~\cite{CASTRADES:2021, RADES:2024}, RBF~\cite{DePanfilis:1987dk, Wuensch:1989sa}, TASEH~\cite{TASEH:2022} and UF~\cite{Hagmann:1990tj, HAGMANN1996209} (adapted from \cite{cajohare:github}). Bottom: Zoom in on results from HAYSTAC (left) \PII and (right) \PI. The dashed red line denotes the joint \SI{90}{\percent} aggregate exclusion level of \PIIJointAgg for all of \PII.  Marked in gray are regions removed from the exclusion due to RFI contamination described in the text.  The QCD axion model band is shown in yellow~\cite{Diluzio2020modelband}, with the benchmark KSVZ and DFSZ models~\cite{kim1979KSVZ,shifman1980KSVZ2,dine1981DFSZ,zhit1980DFSZ2} shown as black dashed lines.}
    \label{fig:exclusion}
\end{figure*}

\begin{table}[]
\begin{tabular}{c|cccc}
    $\nu$ {[}GHz{]} & Persist          & 0-Field        &   Amb.~Veto  \\ \hline
    4.676364     & \checkmark           & X              & X \\
    4.625081     & X                   & X              & X  \\
    4.625005     & X                    & \checkmark     & X   \\
    4.624968     & X                    & \checkmark     & X          \\
    4.624940     & X                     & \checkmark     & X         \\
    4.603546     & X                     & X              & X  \\
    4.603480     & X                   & X              & X   \\
    4.437575     &\checkmark            & X              & X  \\
    4.437503     &\checkmark         & X              & X  \\
    4.316643     & X                    & \checkmark     & X      \\
    4.306543     & X                 & \checkmark     & X      \\
    4.306533     & X                  & X              & X      \\
\end{tabular}    
\caption{Summary table of the \numRFI RFI detected in \PIIcd showing the three pass~(\checkmark)/fail~(X) checks  required to be considered an axion. This includes persistence upon further observation (Persist), failure to persist when the magnetic field is absent~(0-Field), and lack of clear correlation to an ambient rf signal~(Amb.~Veto). Each candidate failed at least one check and was eliminated as an axion signal.}
\label{tab:RFIs}
\end{table}

Using the grand spectrum produced from an initial scan, potential candidates are identified as excesses $\geq3.468\sigma$, corresponding to a $10\%$ two-scan false negative rate for a 5.1$\sigma$ target significance in the frequentist framework used in~\cite{zhong2018results,brubaker2017first,brubaker2017analysis}. Any candidate found above this threshold was further interrogated in rescans to test for persistence, as expected for an axion signal. This procedure identified 98 candidates for \PIIc and 120 for \PIId. While most of these candidates are likely the result of random fluctuations in the noise, three distinct populations of non-Gaussian noise were identified as summarized in Table~\ref{tab:summary}.

The first group were the six synthetic axion SIs described in the previous section. Each SI candidate was identified above the threshold at the injected frequency and data in the \SI{10}{kHz}~($\sim$2$\delta\nu_a$) window around each candidate was cut and later filled in during rescans.  In addition, the shape and scaling with cavity detuning were examined for each of the injected signals to validate the hardware and analysis chains.  An example of this validation is shown in Fig.~\ref{fig:SI} for one of the injections.  As can be seen, the shape in the grand spectrum after combining all contributing PSDs is well matched to that of the injected signal, and the amplitude of the signal in each of the PSDs closely follows the expected scaling from the SNR model detailed in~\cite{HAYSTAC_2023MJ}. Next was a group of candidates, also seen in \PIIab, presenting as large ($\sim$\SI{40}{kHz}) power ``deficits" exceeding the $5.1\sigma$ target significance but in the negative direction. While their source is unknown, no such candidate has been found to repeat upon rescan. Data cuts of \SI{200}{kHz} were conservatively applied to remove data around each candidate and the gaps were later filled in upon rescan. The final group are a set of large but narrow~($\sim$\SI{1}{kHz}) excesses which are clearly visible above the target significance in both the grand spectrum and the PSD from multiple tuning steps, with their clear signature prompting suspicion of rf interference~(RFI) from environmental sources. Upon probing each candidate a total of three times with field on at \SI{8}{T}, only three candidates persisted in all scans as summarized in Table \ref{tab:RFIs}. Each candidate was also scanned at zero-field, with the three persistent candidates all still visible in the spectrum.  The five candidates which did not persist at zero-field also failed to persist during the three rescans taken at \SI{8}{T}, likely indicating a change in the source or in the coupling to the detector rather than an axion signal. The final step in confirming these excesses as RFI was to check against an ambient rf detector consisting of a simple antenna. All \numRFI candidates had a clear correlation to an ambient signal within $\delta_{\nu_{a}}$$\sim$~\SI{5}{kHz}, ruling out axions and other sources such as Dark Photons~\cite{darkphoton_bandbook,darkphoton_sumita}. The exact origin of these signals is unknown, but our detection band is between 4--5~GHz, a popular band for communication~\cite{wifi}.  To remove them from the grand spectrum, data in the \SI{10}{kHz} window around each RFI were cut, amounting to a $\sim$0.02$\%$ loss in total frequency coverage in \PII.

After completing rescans, all candidates either failed to persist or were ruled out through the cross-checks described above. Given the absence of an axion signal, an exclusion limit on $|g_\gamma|$ is set using the Bayesian framework outlined in~\cite{palken2020improved} giving both a \SI{10}{\percent} prior update contour for each scanned frequency along with an aggregate exclusion at the \SI{90}{\percent} level over the entire range.  Aggregated separately, couplings $|g_\gamma|\geq$~\PIIcJointAgg are excluded between 18.71--19.46~$\mu$eV in \PIIc and couplings $|g_\gamma|\geq$~\PIIdJointAgg are excluded between 17.28--18.44~$\mu$eV in \PIId (excluding the mode crossings at 17.89--18.08~$\mu$eV and 18.13--18.18~$\mu$eV and the \numRFI RFI sources). In addition, the individual subphases can be combined to find a joint aggregate exclusion level of \PIIJointAgg over the entire range covered in \PII. The uncertainty on the reported  $|g_\gamma|$ is estimated from the individual parameter uncertainty described in \cite{HAYSTAC_2023MJ} and found to be 9.2$\%$. These results, along with previous HAYSTAC results, are plotted in Fig.~\ref{fig:exclusion}.  

\textit{Conclusion and Outlook}---The results presented in this Letter show the combined search of HAYSTAC's \PII operation, which covers \PIIFreqCovMHz of parameter space in the extended QCD model band including \PIIcdFreqCovMHz of newly explored parameter space. This result showcases the SSR's capability to operate effectively across a wide range of parameter space. Future operations will focus on expanding HAYSTAC to higher frequencies with a new set of JPAs, a new cavity design~\cite{Simanovskaia:2020hox}, techniques to mitigate squeezing degradation related to vibration, and a new readout involving state swapping and two-mode squeezing to further speed up the scan rate~\cite{Wurtz_2021,Jiang_2023}.   

\hfill \break
%\begin{acknowledgments}
%TC:ignore

\textit{Acknowledgments}---HAYSTAC is supported by the National Science Foundation under grants PHY-1701396, PHY-1607223, PHY-1734006, PHY-2011357, PHY-2309631, PHY-2209556, the Heising-Simons Foundation under grants 2014-0904 and 2016-044, and the Sloan Foundation under grant FG-2022-19263 141275. We thank Kyle Thatcher and Calvin Schwadron for their work on the design and fabrication of the SSR mechanical components, Felix Vietmeyer for his work on the room temperature electronics, and Steven Burrows for his graphical design work. We thank Vincent Bernardo and the J.~W.~Gibbs Professional Shop as well as Craig Miller and Dave Johnson for their assistance with fabricating the system’s mechanical components. We also wish to thank the Cory Hall Machine shop at UC Berkeley and the efforts of Sergio Velazquez for fabricating several of the prototype components tested here. We thank Dr.~Matthias Buehler of low-T Solutions for cryogenics advice. Finally, we thank the Yale Wright laboratory for housing the experiment and providing computing and facilities support.    
%TC:endignore
%\end{acknowledgments}

\appendix

%\end{document}

\bibliography{apssamp,refs_haloscopes}% Produces the bibliography via BibTeX.
\end{document}